This page is a placeholder inserted by the administrators so that the system could produce PDF for this paper. The real paper is Walker.ps, which follows this page if you're looking at the PDF.

# Propagation Speed of Longitudinally Oscillating Gravitational and Electrical Fields

*William D. Walker   and   J. Dual*

(General relativistic analysis performed in collaboration with T. Chen)

Institute of Mechanics, Swiss Federal Institute of Technology, 8092 Zurich, Switzerland
E-mail:  walker@ifm.mavt.ethz.ch,   dual@ifm.mavt.ethz.ch

For several years, the authors have been investigating the possibility of developing a laboratory experiment capable of measuring the speed of gravitational interaction. During the 1950s - 1960s, several researchers proposed that it might be possible to longitudinally vibrate a mass near another mass and to detect the resultant gravitationally induced longitudinal vibration. The phase speed could then be determined from the oscillation frequency, the separation distance between the masses, and the measurement of the phase difference of the vibration between the two masses. These early researchers had assumed that the phase speed of gravity was equal to the speed of light. The phase shift expected for a typical experimental set-up was on the order of 1 microdegree. Because of the limited technology at the time, no gravitational experiments were performed. In 1963 R. P. Feynman published a general physics book in which he analysed the electric field of an oscillating charge. Feynman's conclusion was that the oscillating field propagates nearly instantaneously along the axis of vibration, much faster than the speed of light. Because of the similarity of the analogous oscillating mass problem, the physics community has since concluded that the phase speed of both a longitudinally oscillating gravitational field and a longitudinally oscillating electrical field are too fast to measure with a near-field laboratory experiment.

Feynman's analysis is not valid in the near field and is therefore inconclusive for a near-field laboratory experiment. An analysis of the electrical field produced along the axis of vibration of an oscillating charge, valid in the near field, is presented. The solution indicates that the phase speed of the longitudinally oscillating electric field next to the vibrating charge ($r \approx d_o$) is nearly infinite, and rapidly decays to the form $c^3/(\omega^2 r^2)$ in the far field ($r >> d_o$), in which $d_o$ is the vibration amplitude of the charge, c is the speed of light, $\omega$ is the frequency of vibration, and r is the distance from the centre of the oscillating charge. A preliminary calculation of the gravitational field produced by the analogous oscillating mass problem is also presented. The solution indicates that the phase speed of a longitudinally oscillating gravitational field is equal to or larger than order $c^2r/(\omega d_o^2)$. Both of these results indicate that the phase speeds of a longitudinally oscillating electrical and gravitational field are too large to be measurable with a laboratory experiment.

The possibility of measuring the group speed of a longitudinally oscillating gravitational field, which is commonly thought to be equal to the speed of light, is now being considered. The basic idea is to amplitude-modulate the longitudinal vibration of a mass and to measure the resultant longitudinal vibration of a nearby mass due to gravitational interaction. The modulation signal can be extracted using a diode detector and the group speed can then be determined from the oscillation frequency, the separation distance between the masses, and the measurement of the phase shift of the modulation signal. If the group speed is equal to the speed of light, phase shifts on the order of 1 microdegree could be generated with a typical experimental set-up. It is presently unknown if the phase stability of an experimental system can be controlled to this accuracy over the measurement time. A bending gravitationally interacting system that is capable of generating a nanometer gravitationally induced vibration amplitude, which is 4 orders of magnitude larger than previously achieved with other systems, has been developed and tested. The resultant gravitationally induced vibration is in good agreement with Newtonian calculations. In addition, a phase measurement system that is capable of measuring a 100 nanodegree phase shift and a 5 nanovolt change in amplitude, which is 5 orders of magnitude more sensitive than commercial lock-in amplifiers, has also been developed and tested.

# Theoretical Analysis of the Phase Speed of a Longitudinally Oscillating Electrical Field

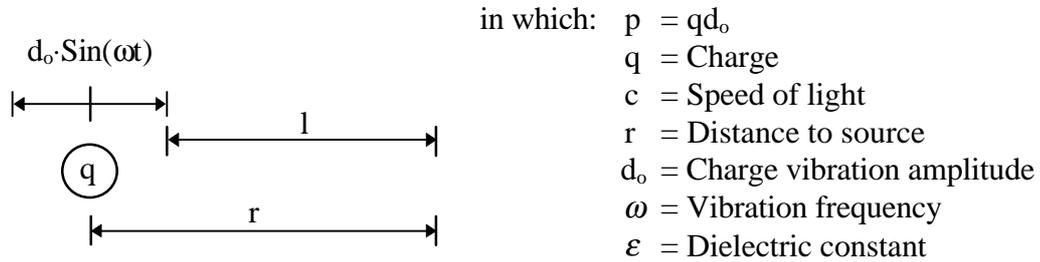

in which:  p = qd$_o$
q = Charge
c = Speed of light
r = Distance to source
d$_o$ = Charge vibration amplitude
$\omega$ = Vibration frequency
$\varepsilon$ = Dielectric constant

**Figure 1**: Vibrating charge model used to calculate the phase speed of an oscillating electric field along the axis of vibration.

### *R. P. Feynman Solution[a]:*

(Used multipole analysis. Valid only in far field i.e. r >> d$_o$)

$$E_{axis_{AC}} \underset{r>>d_o}{=} -\frac{1}{2\pi\varepsilon r^3}\left[p\left(t-\frac{r}{c}\right)+\frac{r}{c}\dot{p}\left(t-\frac{r}{c}\right)\right]$$

Although not calculated by Feynman, the following conclusions can be deduced from this result:

$$E_{Axis_{AC}} \underset{r>>d_o}{=} -\frac{qd_o}{2\pi\varepsilon r^3}\sqrt{1+\frac{r^2\omega^2}{c^2}}\left[Sin(\omega t+\theta)\right]$$

in which:

$$\theta \underset{r>>d_o}{=} Tan^{-1}\left[\frac{\frac{\omega r}{c}Cos\left(\frac{\omega r}{c}\right)-Sin\left(\frac{\omega r}{c}\right)}{Cos\left(\frac{\omega r}{c}\right)+\frac{\omega r}{c}Sin\left(\frac{\omega r}{c}\right)}\right]$$

Taylor expanding the result for $r < \frac{c}{\omega}$ yields the following series:

$$\theta \underset{d_o<<r<<\frac{c}{\omega}}{=} -\frac{1}{3}\left(\frac{\omega^3 r^3}{c^3}\right)+\frac{1}{5}\left(\frac{\omega^5 r^5}{c^5}\right)-\frac{1}{7}\left(\frac{\omega^7 r^7}{c^7}\right)+\cdots \underset{d_o<<r\leq\frac{c}{3\omega}}{\approx} -\frac{1}{3}\left(\frac{\omega^3 r^3}{c^3}\right) \quad \text{10% accuracy}$$

---

[a] R. P. Feynman, Feynman lectures in physics, Addison - Wesley Pub. Co., Vol. 2, Ch. 21, (1964).

The phase speed ($c_{ph}$) of an oscillating field of the form: $Sin(\omega t - kr)$, in which $k = k(\omega, r)$, can be determined by setting the phase part of the field to zero, differentiating the resultant equation, and solving for $\partial r / \partial t$:

$$\frac{\partial}{\partial t}(\omega t - kr) = 0 \qquad \therefore \omega - k\frac{\partial r}{\partial t} - r\frac{\partial k}{\partial r}\frac{\partial r}{\partial t} = 0 \qquad \therefore c_{ph} = \frac{\partial r}{\partial t} = \frac{\omega}{k + r\frac{\partial k}{\partial r}}$$

Differentiating $\theta \equiv -kr$ with respect to r yields: $\quad \dfrac{\partial \theta}{\partial r} = -k - r\dfrac{\partial k}{\partial r}$

Combining these results yields: $\quad c_{ph} = -\omega \Big/ \dfrac{\partial \theta}{\partial r} \quad$ in which, $\theta$ is the phase shift

The phase speed of the longitudinally oscillating electrical field is calculated to be:

$$c_{ph} = -\omega \Big/ \frac{\partial \theta}{\partial r} \underset{r \gg d_o}{=} c + \frac{c^3}{\omega^2 r^2} \underset{r \to \infty}{\lim} = c$$

$$\therefore c_{ph} \underset{d_o \ll r \leq \frac{c}{3\omega}}{\approx} \frac{c^3}{\omega^2 r^2} \qquad 10\% \text{ accuracy}$$

$$\therefore c_{ph} \underset{r \geq \frac{3c}{\omega}}{\approx} c \qquad 10\% \text{ accuracy}$$

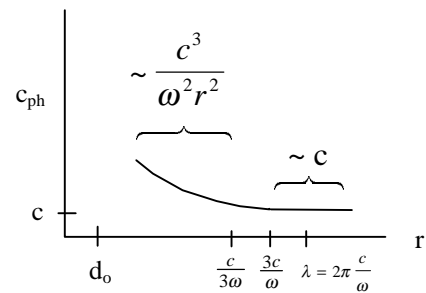

### *Analysis using the Liénard-Wiechert potentials[b] :*

(Valid in near field, ref. Figure 1)

$$\overline{E} = -\nabla \varphi - \frac{d\overline{A}}{dt}$$

$$\vartheta = \left[\frac{K_o}{\left(1 - \frac{\overline{v}}{c}\cdot \overline{n}\right)l}\right]_{ret} \qquad \overline{A} = \left[\frac{K_o \dfrac{\overline{v}}{c}}{\left(1 - \frac{\overline{v}}{c}\cdot \overline{n}\right)l}\right]_{ret}$$

in which:
$\overline{E}$ = Electric field
$\varphi$ = Scalar potential
$\overline{A}$ = Vector potential
$\overline{v}$ = Velocity of the moving charge
$\omega$ = Vibration frequency
$\overline{n}$ = Unit vector from source to observation point
ret (or ') = Quantity in brackets to be evaluated at the retarded time (Tr)

$$K_o = -\frac{q}{4\pi\varepsilon}$$

---
[b] J. Jackson, Classical Electrodynamics, John Wiley &Sons, (1975).

$$\therefore \bar{E} = K_o \left[ \frac{\left(\bar{n} - \frac{\bar{v}}{c}\right)\left[1 - \left(\frac{\bar{v}}{c}\right)^2\right]}{\left(1 - \frac{\bar{v}}{c} \cdot \bar{n}\right)^3 l^2} \right]_{ret} + \text{far-field transverse radiation term}$$

$$\therefore E_{Axis} = K_o \left[ \frac{1 - \left(\frac{v'}{c}\right)^2}{(l')^2 \left(1 - \frac{v'}{c}\right)^2} \right]$$

in which:

$$x = d_o \mathrm{Sin}(\omega t) \qquad v = \frac{dx}{dt} = d_o \mathrm{Cos}(\omega t) \qquad \therefore \frac{v}{c} = \beta \mathrm{Cos}(\omega t) \qquad \beta = \frac{\omega d_o}{c}$$

$$\therefore \frac{v'}{c} = \beta \mathrm{Cos}(\omega T_r) \qquad l = r - d_o \mathrm{Sin}(\omega t) \qquad \therefore l' = r - d_o \mathrm{Sin}(\omega T_r) \qquad T_r = t - \frac{l'}{c}$$

Fourier transforming the result yields:

$$E_{Axis} \underset{d_o \leq r}{=} \frac{-q_{Tx}}{4\pi\varepsilon} \cdot \frac{1}{r^2} \left[1 - \left(\frac{d_o}{r}\right)^2\right]^{-\frac{3}{2}} \left[1 + 2\frac{d_o}{r} \mathrm{Sin}(\omega t + \theta)\right] \left[1 + O\left(\frac{\omega d_o}{c}\right)^2\right] + h.h.$$

in which:

$$\theta \underset{d_o \leq r}{=} \left[1 - \left(\frac{d_o}{r}\right)^2\right]^{\frac{3}{2}} \left[-\frac{1}{3}\left(\frac{\omega^3 r^3}{c^3}\right)\right] + O\left(\frac{\omega d_o}{c}\right)^4 \qquad h.h. = \text{Higher harmonics}$$

The phase speed is calculated to be:

$$c_{ph} \underset{d_o \leq r}{=} -\omega \Big/ \frac{\partial \theta}{\partial r} \underset{d_o \leq r}{=} \underbrace{\left[1 - \left(\frac{d_o}{r}\right)^2\right]^{-\frac{1}{2}} \frac{c^3}{\omega^2 r^2}} \left[1 + O\left(\frac{\omega d_o}{c}\right)^4\right]$$

New near-field term different from Feynman solution

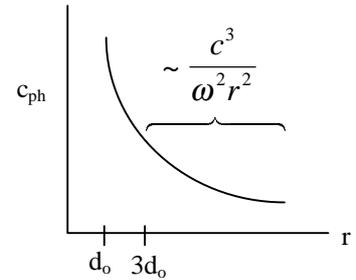

$$\therefore c_{ph} \underset{r \geq 3d_o}{\approx} \frac{c^3}{\omega^2 r^2} \qquad \text{10\% accuracy}$$

## *Conclusion*

The phase speed of a longitudinally oscillating electrical field is much faster than the speed of light and not measurable using modern technology which is only presently capable of measuring phase speeds less than or equal to the speed of light.

# Theoretical Analysis of the Phase Speed of a Longitudinally Oscillating Gravitational Field

*(Analysis performed in collaboration with T. Chen, Institute of Mechanics, ETH, Zürich [c])*

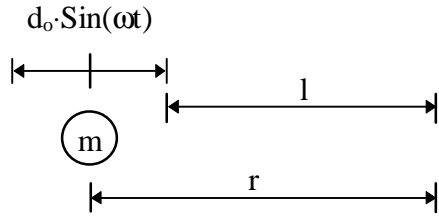

In which:  
$G_{\mu\nu}$ = Einstein tensor  
$T_{\mu\nu}$ = Energy momentum tensor  
$G$ = Gravitational constant  
$c$ = Speed of light  
$\vartheta$ = Gravitational potential  
$\rho$ = Mass density  
$\omega$ = Vibration frequency

**Figure 2**: Vibrating mass model used to calculate the phase speed of an oscillating gravitational field along the axis of vibration.

Using the Einstein relation:

$$G_{\mu\nu} = \frac{8\pi G}{c^4} T_{\mu\nu}$$

Along the axis of vibration, for small masses and low velocities, the Einstein equation reduces to:

$$\left(\frac{1}{c^2}\frac{\partial^2}{\partial^2 t} - \frac{\partial^2}{\partial r^2}\right)\vartheta = -\frac{4\pi G}{c^2} T_{oo}$$

The only nonvanishing term in the energy momentum tensor to order $\beta^2$ is:

$$T_{oo} = \rho c^{2\cdot}\delta[r - d_o Sin(\omega t)]\ [1 + O(\beta^2)] \qquad \text{in which:} \quad \beta = \frac{\omega d_o}{c}$$

Solving the partial differential equation yields:

$$\vartheta = \frac{K_o}{r}\left[\frac{1}{1-\xi Sin\left(\omega t - \frac{\omega l'}{c}\right)} \frac{1}{1-\beta Cos\left(\omega t - \frac{\omega l'}{c}\right)}\right][1 + O(\beta^2)]$$

in which:  $K_o = -mG$,   $\xi = \dfrac{d_o}{r}$,   $l' = r - d_o Sin\left(\omega t - \dfrac{\omega l'}{c}\right)$

---

[c] T. Chen, W. D. Walker, J. Dual, Gravitational near-field interaction, Currently under review by Physical Review

The gravitational field (g) can then be calculated using the relation $g = -\frac{\partial}{\partial r}\vartheta$:

$$g = -\frac{mG}{r^2} \cdot \left[1 - \left(\frac{d_o}{r}\right)^2\right]^{-\frac{3}{2}} \left[1 + 2\frac{d_o}{r} Sin\left(\omega t + O\left(\frac{\omega d_o}{c}\right)^2\right)\right] \left[1 + O\left(\frac{\omega d_o}{c}\right)^2\right] + h.h.$$

The ratio of the phase speed of the gravitational field to the speed of light is calculated to be:

$$\frac{c_{ph}}{c} \geq O\left[\left(\frac{r}{\omega d_o^2}\right) c\right]$$

### *Conclusion*

The phase speed of a longitudinally oscillating gravitational field is much faster than the speed of light and not measurable using modern technology which is only presently capable of measuring phase speeds less than or equal to the speed of light. As in the case of the longitudinally oscillating electric field, calculation of the gravitational field to higher orders in $\beta$ may enable the phase speed to be determined.

## Proposed Experiment to Measure the Group Speed of a Longitudinally Oscillating Gravitational Field

The possibility of measuring the group speed of a longitudinally oscillating gravitational field, which is commonly thought to be equal to the speed of light, is now being considered. The basic idea is to amplitude-modulate the longitudinal vibration of a mass and to measure the resultant longitudinal vibration of a nearby mass due to gravitational interaction. The modulation signal can be extracted using a diode detector and the group speed can then be determined from the oscillation frequency, the separation distance between the masses, and the measurement of the phase shift of the modulation signal. If the group speed is equal to the speed of light, phase shifts on the order of 1 microdegree could be generated with a typical experimental set-up. It is presently unknown if the phase stability of an experimental system can be controlled to this accuracy over the measurement time. Note that the group speed of light has been measured using this type of set-up[d].

---

[d] R. Barr, T. Armstrong, An inexpensive apparatus for measurement of the group velocity of light in transparent media using a modified Helium-Neon laser, Am J. Phys., Vol. 58, No. 11, Nov. (1990).

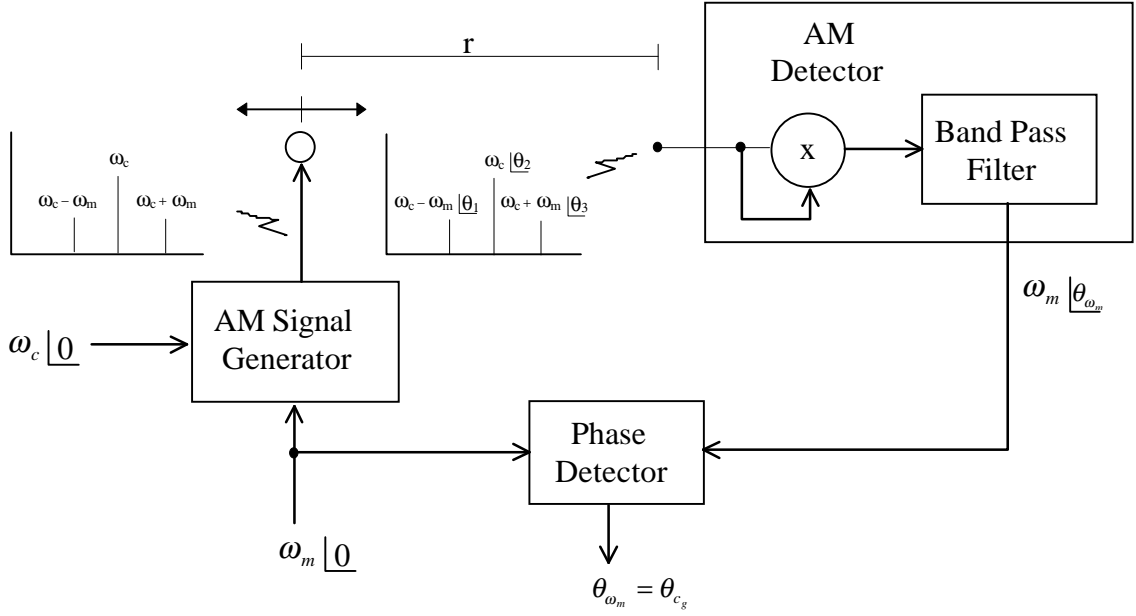

**Figure 3**: Proposed experimental set-up to measure the group speed of a longitudinally oscillating gravitational field.

## *Group Speed Analysis of a Longitudinally Oscillating Electrical Field*

If phase speed of an oscillating field is a function of frequency or space, then the group speed will differ from the phase speed. The group speed of an oscillating field of the form: $Sin(\omega t - kr)$, in which $k = k(\omega, r)$, can be determined by considering 2 Fourier components of a wave group:

$$Sin(\omega_1 t - k_1 r) + Sin(\omega_2 t - k_2 r) = Sin(\Delta\omega t - \Delta k r) Sin(\omega t - kr)$$

in which: $\quad \Delta\omega = \dfrac{\omega_1 - \omega_2}{2}, \quad \Delta k = \dfrac{k_1 - k_2}{2}, \quad \omega = \dfrac{\omega_1 + \omega_2}{2}, \quad k = \dfrac{k_1 + k_2}{2}$

The group speed ($c_g$) can then be determined by setting the phase part of the modulation component of the field to zero, differentiating the resultant equation, and solving for $\partial r / \partial t$:

$$\frac{\partial}{\partial t}(\Delta\omega t - \Delta k r) = 0 \qquad \therefore \Delta\omega - \Delta k \frac{\partial r}{\partial t} - r \frac{\partial \Delta k}{\partial r} \frac{\partial r}{\partial t} = 0 \qquad \therefore c_g = \frac{\partial r}{\partial t} = \frac{\Delta\omega}{\Delta k + r \dfrac{\partial \Delta k}{\partial r}}$$

Differentiating $\Delta\theta \equiv -\Delta k r$ with respect to r yields: $\qquad \dfrac{\partial \Delta\theta}{\partial r} = -\Delta k - r \dfrac{\partial \Delta k}{\partial r}$

Combining these results yields: $\quad c_g = -\Delta\omega \Big/ \dfrac{\partial \Delta\theta}{\partial r} = -\left[\dfrac{\partial}{\partial r}\dfrac{\Delta\theta}{\Delta\omega}\right]^{-1} \underset{\frac{\Delta\theta}{\Delta\omega}\,small}{=\lim} -\left[\dfrac{\partial^2\theta}{\partial r \partial \omega}\right]^{-1}$

The group speed of a longitudinally oscillating electrical field, using the Feynman far-field solution, is calculated to be:

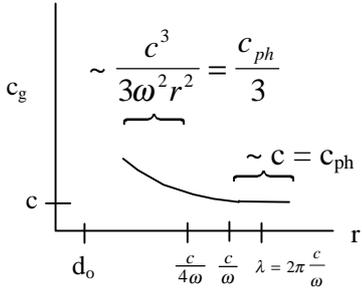

$$c_g = -\left[\frac{\partial^2 \theta}{\partial \omega \partial r}\right]^{-1} \underset{r \gg d_o}{=} \frac{c(c^2 + r^2\omega^2)^2}{3c^2 r^2 \omega^2 + r^4 \omega^4} \underset{r \to \infty}{\lim} = c$$

$$\therefore c_g \underset{d_o \ll r \leq \frac{c}{4\omega}}{\approx} \frac{c^3}{3\omega^2 r^2} = \frac{c_{ph}}{3} \quad \text{10\% accuracy}$$

$$\therefore c_g \underset{r \geq \frac{c}{\omega}}{\approx} c \quad \text{10\% accuracy}$$

Using the Liénard-Wiechert potentials, the group speed of a longitudinally oscillating electrical field in the near field, is calculated to be:

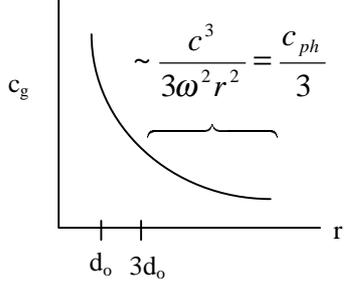

New near-field term different from Feynman solution

$$c_g = -\left[\frac{\partial^2 \theta}{\partial \omega \partial r}\right]^{-1} \underset{d_o \leq r}{=} \left[1 - \left(\frac{d_0}{r}\right)^2\right]^{-\frac{1}{2}} \frac{c^3}{3\omega^2 r^2} \left[1 + O\left(\frac{\omega d_0}{c}\right)^4\right]$$

$$\therefore c_g \underset{r \geq 3d_o}{\approx} \frac{c^3}{3\omega^2 r^2} = \frac{c_{ph}}{3} \quad \text{10\% accuracy}$$

Note that these solutions indicate that group speed in the semi-near-field is faster than the speed of light, which appears to violate causality and should not be possible.

An alternate approach to calculating the group speed is presented which uses amplitude modulation (AM) and demodulation (ref. Figure 3):

The signal out of the AM signal generator is:

$$AM\ Sig = [1 + Sin(\omega_m t)] Sin(\omega_c t)$$

After the electric field has propagated a distance r, the signal out of the multiplier in the AM detector is:

$$AM\ Sig\ Det = \left[\frac{1}{2} Cos[(\omega_c - \omega_m)t + \theta_1] + Sin[\omega_c t + \theta_2] - \frac{1}{2} Cos[(\omega_c + \omega_m)t + \theta_3]\right]^2$$

Inserting the phase solution obtained from the Feynman solution ($d_o << r \leq \frac{c}{3\omega}$):

in which:

$$\theta_1 = \frac{(\omega_c - \omega_m)^3 r^3}{3c^3} \qquad \theta_2 = \frac{(\omega_c)^3 r^3}{3c^3} \qquad \theta_3 = \frac{(\omega_c + \omega_m)^3 r^3}{3c^3}$$

Simplifying the result, the phase of the resultant $\omega_m$ component yields:

$$\theta_{\omega_m}\bigg|_{d_o << r \leq \frac{c}{3\omega}} = \frac{-r^3 \omega_m (3\omega_c^2 + \omega_m^2)}{3c^3}$$

The group speed of the modulated signal can be calculated, yielding the same result as obtained using the classical definition ($d_o << r \leq \frac{c}{3\omega}$):

$$c_g = -\omega_m \bigg/ \frac{\partial \theta_m}{\partial r}\bigg|_{d_0 << r \leq \frac{c}{3\omega}} = \frac{c^3}{r^2(3\omega_c^2 + \omega_m^2)} \underset{\omega_c >> \omega_m}{=} \frac{c^3}{3r^2 \omega_c^2} = \frac{1}{3} c_{ph}$$

## *Conclusion*

The analysis of the group speed of a longitudinally oscillating electrical field is currently inconclusive. The group speed is commonly thought to be equal to the speed of light, but preliminary analysis indicates that the group speed is much faster than light which is not thought possible due to causality violation.

# Experimental Gravitationally Vibrating System

## *Simple Mass Spring Gravitationally Interacting System*

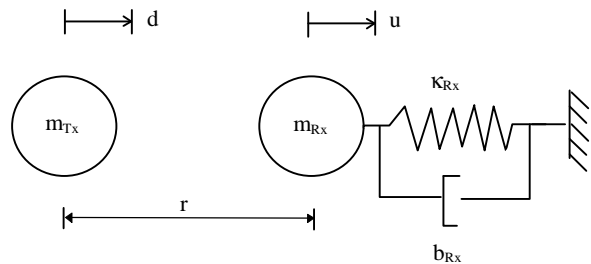

**Figure 4**: Mathematical model of a mass spring gravitationally interacting system.

Inserting a oscillating gravitational force into a damped mass spring partial differential equation yields the following receiver mass vibration amplitude (at resonance):

$$u_{AC\,amp} = \frac{-2GQ_{Rx}m_{Tx}d_o}{\omega_{Tx}^2 r^3}\left[1-\left(\frac{d_o}{r}\right)^2\right]^{-\frac{3}{2}}$$

This result clearly indicates that the receiver's gravitationally induced vibration ($u_{AC}$) is increased by maximising the receiver mass' quality factor ($Q_{Rx}$), the vibrating mass ($m_{Tx}$), and the amplitude of the transmitting mass' vibration ($d_o$). In addition, minimising the vibration frequency ($\omega_{Tx}$) and the distance between the two masses (r) is especially effective.

## New Bending Gravitationally Interacting System

A new bending beam gravitationally interacting system has been built and tested[e]. The system is capable of generating nanometer gravitationally induced vibrations which is 4 orders of magnitude larger than previous systems. The system consists of a brass beam (Tx) which is electromagnetically vibrated at its first mode 40 Hz bending frequency. An adjacent quartz glass beam located 2.5 cm away has been observed to vibrate at the same frequency due to gravitational interaction. The observed vibration is not affected by changing the acoustical , mechanical, or electromagnetic coupling between the two beams. When the spacing between the beams is increased from 2.0 cm to 7.0 cm, a factor of 7.4 reduction in the Rx beam vibration is observed, as predicted from Newtonian theory.

Specifications:
- Tx beam - Brass, 95.9 cm x 11.0 mm x 12.0 mm
  - 1.0 cm beam end, 40 Hz first mode vibration
- Rx beam - Quartz glass, 96.45 cm x 7.6 mm dia., Q ~ 210,000
  - 2.8 nm beam end, 40 Hz first mode vibration experimentally observed
  - 3.8 nm beam end vibration predicted from Newtonian theory

---

[e] W. D. Walker, J. Dual, Experiment to measure the propagation speed of gravitational interaction, Virgo gravity wave conference proceedings, (1996).

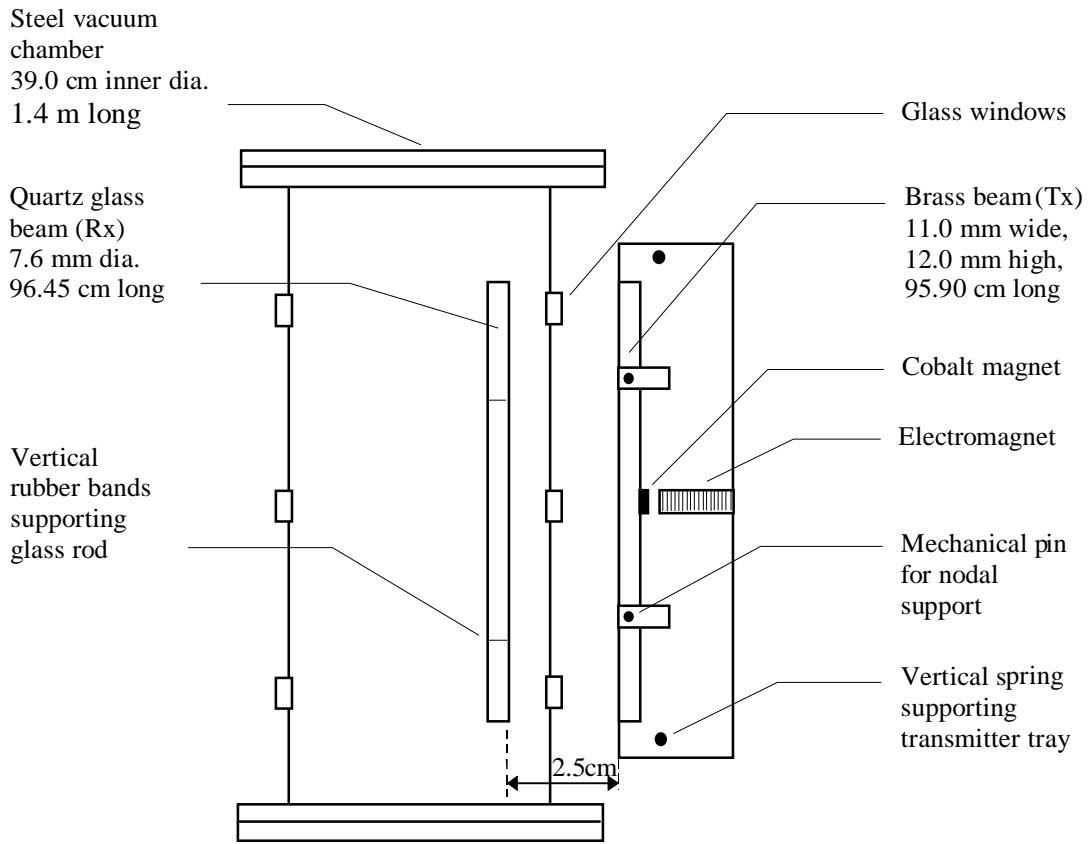

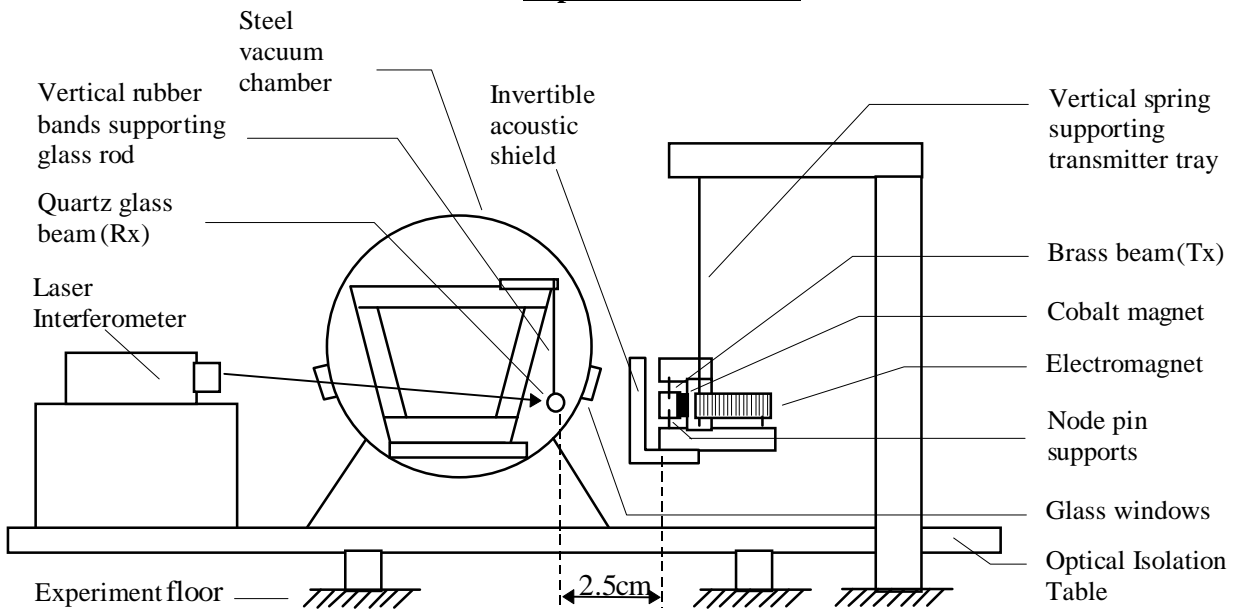

**Figure 5**: Experimental set-up of bending beam gravitationally interacting system.

4...-

# 100 Nanodegree Phase and 5 Nanovolt Amplitude Measurement System

A new differential phase and amplitude measurement system using a commercial lock-in amplifier has been developed and tested[f]. The system is capable of measuring 100 nanodegree phase shifts and 5 nanovolt changes in amplitude, which is 5 orders of magnitude more sensitive than commercial lock-in amplifiers.

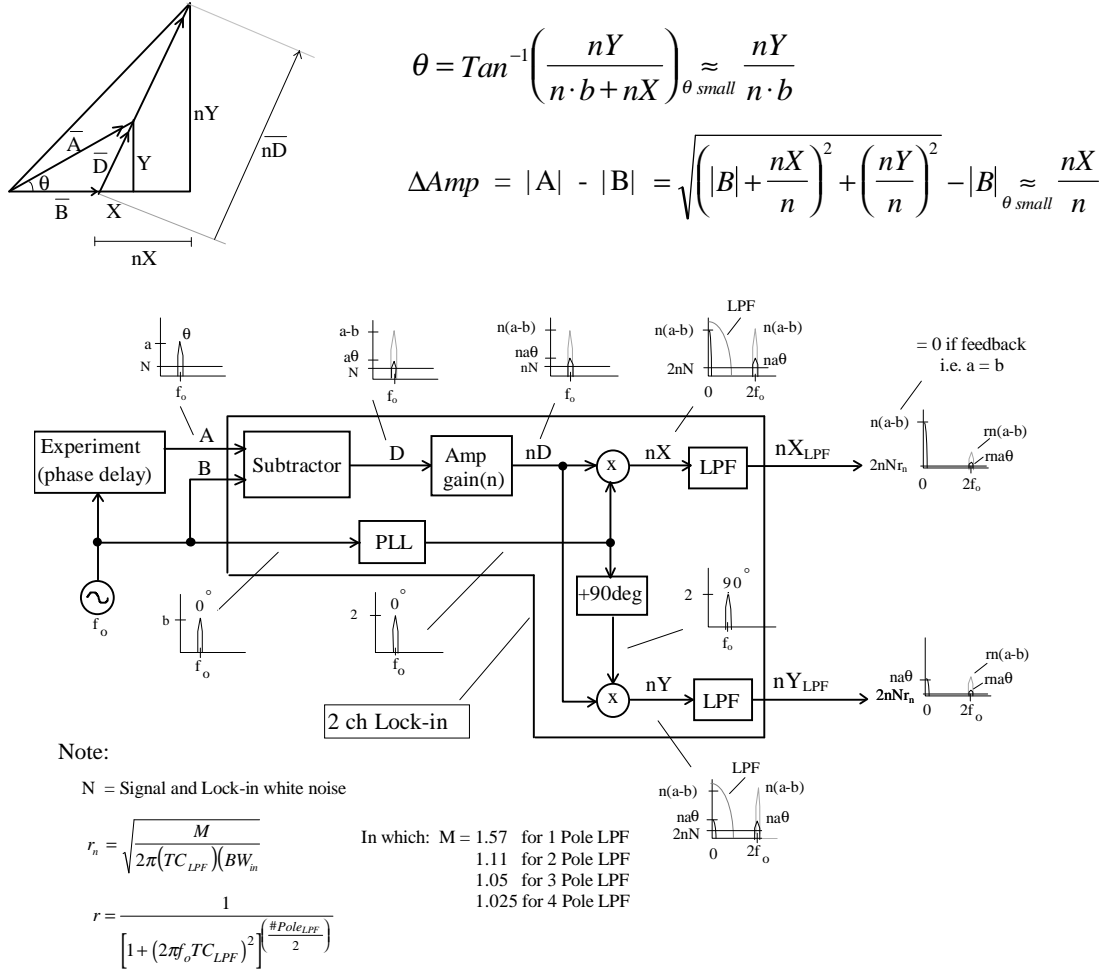

$$\theta = Tan^{-1}\left(\frac{nY}{n\cdot b+nX}\right)_{\theta\,small} \approx \frac{nY}{n\cdot b}$$

$$\Delta Amp = |A| - |B| = \sqrt{\left(|B|+\frac{nX}{n}\right)^2 + \left(\frac{nY}{n}\right)^2} - |B| \Big|_{\theta\,small} \approx \frac{nX}{n}$$

$$r_n = \sqrt{\frac{M}{2\pi(TC_{LPF})(BW_{in})}}$$

$$r = \frac{1}{\left[1+(2\pi f_o TC_{LPF})^2\right]^{\left(\frac{\#Pole_{LPF}}{2}\right)}}$$

In which:  M = 1.57 for 1 Pole LPF
   1.11 for 2 Pole LPF
   1.05 for 3 Pole LPF
   1.025 for 4 Pole LPF

**Figure 6**: Signal processing diagram of differential phase and amplitude measurement.

The maximum sensitivity of this technique is limited by the internal lock-in amplifier noise ($X_{noise}$ and $Y_{noise}$ typically: $5nV/\sqrt{Hz}$) and the bandwidth (BW):

$$\theta_{min} = \frac{180}{\pi}\frac{(Y_{noise}/\sqrt{Hz})\sqrt{BW}}{|A|} \qquad \Delta Amp_{min} = (X_{Noise}/\sqrt{Hz})\sqrt{BW}$$

in which: $\quad BW = \frac{M}{2\pi TC} \qquad$ TC = Time constant of lock-in amplifier

---

[f] W. D. Walker, J. Dual, Experiment to measure the propagation speed of gravitational interaction, Virgo gravity wave conference proceedings, (1996).